\documentstyle[preprint,aps]{revtex}

\tighten
\begin{document}
\draft

\title{\flushleft Once again on the `demythologizing' of the clock-in-the-box 
debate of Bohr and Einstein\footnote
{This article is written by V Hnizdo in his private capacity. No official support
or endorsement by the Centers for Disease Control and Prevention is intended
or should be inferred.}} 
\author{\flushleft V Hnizdo}
\address{\flushleft
National Institute of Occupational Safety and Health,
1095 Willowdale Road, Morgantown, WV 26505, USA}
\address{\flushleft\rm
{\bf Abstract}. The reply of de la Torre, Daleo and Garc\'{\i}a-Mata
[Eur. J. Phys. {\bf 23} (2002) L15--L16] to a criticism of their `demythologizing' 
analysis of the clock-in-the-box debate between Einstein and Bohr is commented on.}

\maketitle

\section*{}
\noindent 
The main point of the reply \cite{TDG2} of de la Torre, Daleo and 
Garc\'{\i}a-Mata (TDG)  
to my criticism \cite{VH} of their `demythologizing' analysis \cite{TDG} of the 
arguments of Bohr and Einstein concerning the famous clock-in-the-box
gedanken experiment is that the criticism suffers from
the `same mistakes' as those which TDG assert have been committed by Bohr:
`the confusion of quantum indeterminacies with experimental uncertainties'.
TDG point out that the meaning of the `quantum indeterminacy' $\Delta A$
of an observable $A$ in a state $\psi$ is only given by
$\Delta A=[\langle\psi|A^2|\psi\rangle-\langle\psi|A|\psi\rangle^2]^{1/2}$,
which is a quantity that is essentially different from the `experimental
uncertainty' $\delta A$ in a measurement of $A$, and that while the lack of
appreciation of this difference could have been tolerated during the early
days of quantum mechanics, there is no excuse for making the mistake of
confusing those two quantities today. In the present note, I comment on
this and the other points of the reply of TDG.

If the experimental uncertainty $\delta A$ refers to 
a measurement of the observable $A$ done
on a state in which $A$ has an indeterminacy $\Delta A$, then,
of course, there is no prescribed relation between $\delta A$ and 
$\Delta A$.
A state in quantum mechanics, in general a mixed state describable by 
a density matrix $\rho$, always can be defined so
that the indeterminacy  $\Delta A$ of a continuous-spectrum observable $A$,
defined by $(\Delta A)^2={\rm Tr}(\rho A^2)-[{\rm Tr}(\rho A)]^2$, 
takes in this state an arbitrarily given value, and, in principle,      
the measurement of $A$ in this state can be done with an 
arbitrary uncertainty $\delta A$. 
But one thing is to define a state in a formalism,  and another
is the physical realization of a such a state. In the standard
interpretation of quantum mechanics (the essential features of which
are due to Bohr himself), a physical state to be described by the 
formalism arises as the result of a physical procedure
that is describable in purely classical terms; such a procedure
is usually also called a `measurement'. Measurement in
this role is then a state `preparation'.  
A state preparation's experimental uncertainties, the analysis and assignment 
of which may require elements of quantum physics, determine the 
indeterminacies in the state prepared.
Thus, for example, the simultaneous measurement
of a macroscopic harmonic oscillator's position and momentum  
with experimental uncertainties $\delta q$ and $\delta p$ would
prepare a mixed (as opposed to pure) harmonic-oscillator state in which
the position and momentum indeterminacies are
$\Delta q\approx \delta q$ and $\Delta p\approx\delta p$. There would be no
harmony between the physical possibilities of measurement and
the quantum-mechanical formalism if a preparation with 
$\delta p\,\delta q\ll\hbar$, where $\delta p$ and $\delta q$
are the experimental uncertainties in two conjugate quantities $p$ and $q$,   
was physically possible.
This is the reason why Einstein tried to devise experimental procedures
using which one could prepare, at least in principle, states in which 
two conjugate quantities would be determined arbitrarily sharply,
and why Bohr considered it so important to demonstrate that
any such preparation was impossible.

The balancing procedure of the clock-in-the-box experiment is
a preparation of the box in a state in which  position and momentum
indeterminacies $\Delta q$ and $\Delta p$
match the experimental uncertainties $\delta q$
and $\delta p$ of the preparation. There is no `real' quantum state,
such as some stationary, or coherent, pure state, in which
the macroscopic box `exists' independently of measurement.
Once the simple and basic point of the dual role of measurement
as a state preparation and a state testing is appreciated, there should be
no confusion concerning `uncertainties' and `indeterminacies'
(or `accuracies' and `latitudes' in Bohr's terminology).

TDG made the point that an algebraic manipulation of a given set
of equations and inequalities does not by itsef constitute a correct
derivation of a physically meaningful statement. But 
they produced by just such a manipulation their counterexample
to Bohr's inequality $\Delta p<\Delta m\,g T$ for the uncertainty/indeterminacy
 $\Delta p$
in the box's momentum  after its mass has been measured to an accuracy
$\Delta m$ in a balancing procedure taking a time $T$.
They identify the total bound-state energy
$\frac{1}{2}mv^2+\frac{1}{2}k q^2$ of an oscillator, which is
the sum of its kinetic and elastic potential energies, with
the  rest-mass energy $mc^2$ of the box itself. In their reply, 
TDG defend this by observing that the elastic potential energy 
can be neglected as negligible, which is, of course, true 
with respect to the rest-mass energy $mc^2$ of the box,
but the point here is that their identification is
$mc^2=\frac{1}{2}mv^2+\frac{1}{2}k q^2$,
which implies an absurd mean square velocity 
$\overline{v^2}\sim c^2$ as 
$\frac{1}{2}mv^2+\frac{1}{2}k q^2\sim m\overline{v^2}$ .
The mass the balancing procedure is designed to measure is 
the rest mass $m$  of the box, not the mass equivalent of the
sum of the kinetic and elastic potential energies of the box.
(TDG betrayed a confusion on this point already when they talked
about the need to prevent a dissipation of the kinetic energy
of the box to its environment;
in fact, precisely such a dissipation is desirable, and only a transfer of the
kinetic energy
to the internal energy of the box and hence to its rest mass is not.)
Of course, one may write down a coherent harmonic-oscillator state
$\psi_{\alpha}$ such that its energy indeterminacy 
$\Delta E=\hbar\omega|\alpha|$ is smaller than any arbitrarily given
value by choosing a sufficiently small $|\alpha|$, while the momentum
indeterminacy stays fixed at $\Delta p=(\frac{1}{2}\hbar m\omega)^{1/2}$.
But this $\Delta E$ is the indeterminacy of the energy 
$E=\frac{1}{2}mv^2+\frac{1}{2}k q^2$,
and not of the rest-mass energy $mc^2$ of the box.
In the example of the ground stationary harmonic-oscillator state, 
given by equation (6) of \cite{VH},
one has $\Delta E=\hbar\omega|\alpha|=0$ as $\alpha=0$,
and $\Delta p=(\frac{1}{2}\hbar m\omega)^{1/2}\approx\Delta m\,g\tau/2\pi$,
where $\Delta m$ is correctly
the accuracy of the rest mass of the oscillator as it is measured
by the balancing. This is consistent with Bohr's inequality
$\Delta p<\Delta m\,g T$ because the oscillator period $\tau$ is, very 
conservatively,
a lower bound on the time $T$ needed to prepare such a state by the
balancing procedure. In fact, to prepare a {\it macroscopic} object like 
the box in a {\it pure} quantum state may require 
a time $T$ that is by many orders of magnitude greater than the
time that can be allocated  to any meaningful measurement procedure, 
but that is not an issue for Bohr's analysis as there is no requirement there
that the balancing must prepare a pure state.
The time $T$ needed to prepare the box in a state in which the balancing
measures its rest mass to a given accuracy $\Delta m$ cannot be ignored.
The imagining of TDG of a balancing `quality test' that could `collapse'
a macroscopic box hanging on a spring balance into a state
in which $\Delta p>\Delta m\,g T$ is a totally
unfounded fantasy even if one would allow that the state in question
did not have to be a pure one---if $T $ is the time the test takes and 
$\Delta m$ is the test's tolerance on the box's rest mass $m$. 
In their reply, TDG bring up again a common point of several criticisms
of Bohr's analysis of the clock-in-the-box experiment, namely
that a use of the red-shift formula cannot be a `legal' part of 
an argument concerning quantum mechanics because the validity of the latter
should not depend on the correctness of such a disparate theory as 
general relativity. This point was not the focus of the 
TDG analysis, and thus it was not broached in my criticism.
Here I would like to remark only that it can be shown very simply that the 
red-shift formula 
is a necessary consequence of an assumption that is basic to
the whole idea of the clock-in-the-box experiment, 
namely that energy has weight, independently of
the theories of special and general relativity,
and the principle of equivalence in particular \cite{GU}. 
Also, the question whether mass is a parameter or a quantum-mechanical
observable with a Hermitian operator is of no relevance to the fact
that mass can be measured in a macroscopic balancing procedure. 
In this connection one may note that in fact time
is a parameter in the formalism of quantum mechanics, but this
circumstance does not prevent the formulation
of a meaningful time--energy uncertainty relation---it just means
that, unlike the momentum--position uncertainty relations,
such a relation cannot be straightforwardly derived from the formalism.

In my opinion, the reply of TDG has not properly addressed any of 
the points of my criticism of their `demythologizing'
of the clock-in-the-box debate of Bohr and Einstein.

\end{document}